# The Fire We Share

## From Scars to Seeds: Reimagining Fire Data as Interactive Memory

Chen Wang
California State University, Fullerton
cwang@fullerton.edu

Mengtan Lin
California State University, Fullerton
mengtanlin@csu.fullerton.edu

## Abstract

**The Fire We Share** proposes a care-centered, consequence-aware visualization framework for engaging with wildfire data not as static metrics, but as living archives of ecological and social entanglement [4,5].

By combining plants-inspired data forms, event-based mapping, and narrative layering, the project foregrounds fire as a shared temporal condition—one that cuts across natural cycles and human systems.

Rather than simplifying wildfire data into digestible visuals, The Fire We Share reimagines it as a textured, wounded archive—embodied, relational, and radically ethical.

## Introduction

*The Fire We Share* challenges the myth of neutrality in data visualization. Wildfires are not isolated "natural disasters"—they are co-authored by overlapping human systems (urban sprawl, outdated power infrastructure, failed suppression policy) and natural cycles (drought, ignition ecology, and regenerative burning) [1]. In California, where 94% of recent fires threatening homes are human-caused [6], the line between environment and accountability blurs.

This project transforms wildfire data into a living archive—a structure for collective memory and ecological warning. Using forms inspired by tree rings and fire-activated pinecones, it encodes each fire event not merely through metrics (acres burned, structures lost), but through visual ruptures, broken symmetry, and narrative entanglement.

The pinecone becomes more than metaphor. It is reimagined here as a ritual interface: each scale a tactile site of loss and resilience, each deformation a signal of consequence. Viewers engage with it as a layered story—through 3D form, projection, or AR—touching protruding scales to access maps, memories, and statistics. This interface does not smooth data; it scars it.

Rather than aiming for closure or resolution, The Fire We Share opens a space for multi-perspectival witnessing. It invites viewers not only to see—but to feel, question, and co-grieve.

Like a scorched pinecone that must burn to release its seeds, this visualization is both archive and catalyst. It holds space not only for information, but for reflection, contradiction, and care—offering not answers, but presence. In this light, visualization becomes not just a representational act, but a relational one.

*collective care*

This series consist of three parts:

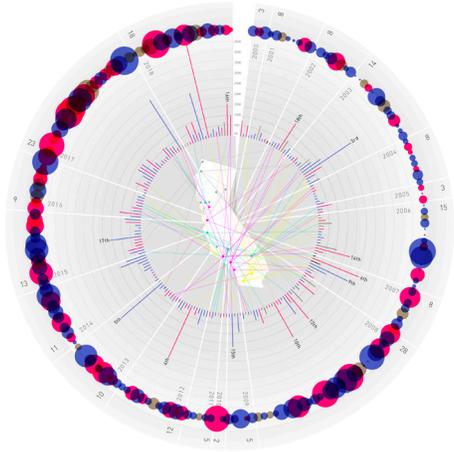

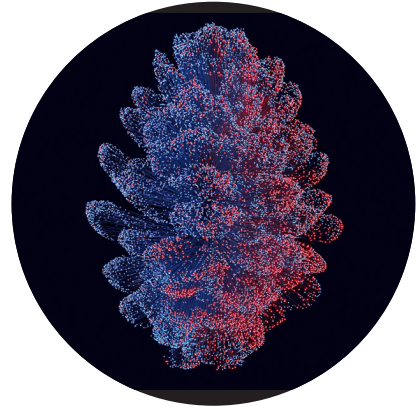

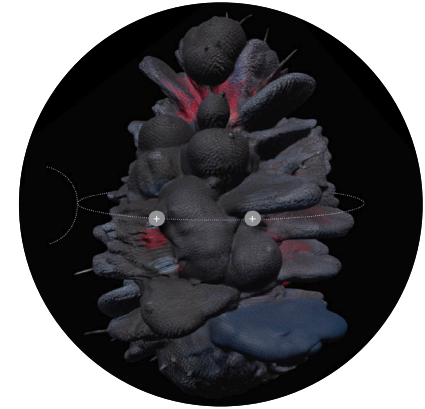

### Part 01: Tree Rings

**Fractured time and layered memory**

Fire data is transformed into temporal scars—ring-like structures that encode ignition causes, ecological impact, and policy events. Each rupture marks not just a moment, but a system's failure and memory.

### Part 02: Pine Cone

**Ritual Interface and Narrative Vessel**

Each scale becomes a point of ignition. Through embodied gestures—touch, rotation, breath—viewers spark stories of loss, resilience, and care. The cone transforms into a ritual vessel for shared witnessing.

### Part 03 — Living Archive

**Layered Storytelling and Shared Voices**

Narratives unfold through time, emotion, and user interaction. Visitors access embedded testimonies and contribute their own voices via a collective upload portal—turning the archive into a polyphonic, co-created memory system.

**Part 01: TREE RINGS**
**Rupture, Memory, and Ecological Time**

Tree rings are not merely markers of age—they are ecological memory systems, inscribed with rupture, rhythm, and regeneration. In this project, the radial form becomes a visual and symbolic framework for mapping wildfire events not as isolated incidents, but as layered entanglements of data, policy, and lived experience.

Each ring encodes more than just ignition cause, fire scale, and geographic location. It also embeds what remains unspoken: policy shifts, silenced communities, funding disparities, and ecological trauma [2,3].

These radial scars reflect how time is not only measured, but wounded. The missing rings, dark burns, and fractured lines bcome forms of embodied data—carried not in spreadsheets, but in scorched wood and the silence that follows.

Through this structure, time becomes a witness. The archive grows not linearly, but concentrically, allowing us to revisit, reinterpret, and re-layer the past. This is not a diagram. It is a ritual cartography—a circle of fire, loss, and co-witnessing [4,5].

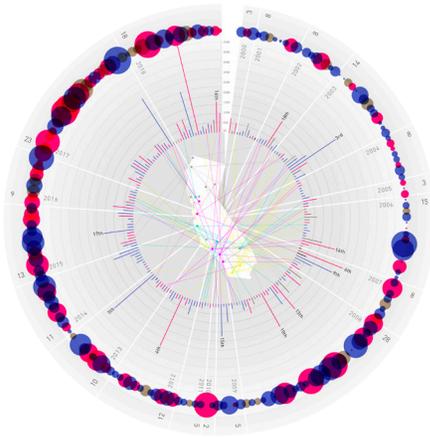

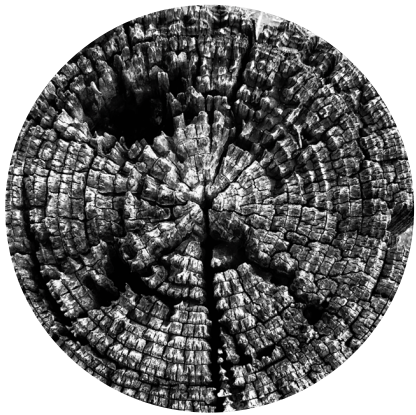

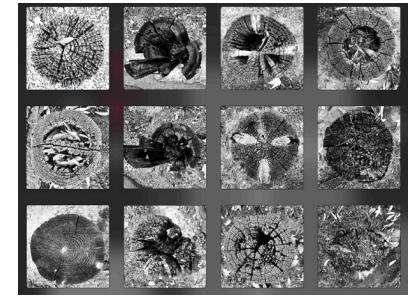

Wildfire data table (2000–2003), compiled from CAL FIRE and USFS incident archives. Key variables include ignition cause, acres burned, vegetation type, and impact metrics. Selected events are later encoded in the tree ring visualizations[1].

Field documentation of post-fire tree cross-sections, photographed in Southern and Central California burn sites. Each scarred ring functions as a physical memory archive—indexing not only heat and duration, but also ecological absence, temporal rupture, and patterns of regrowth.

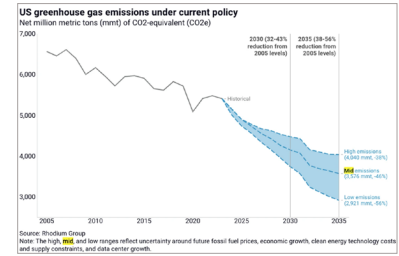

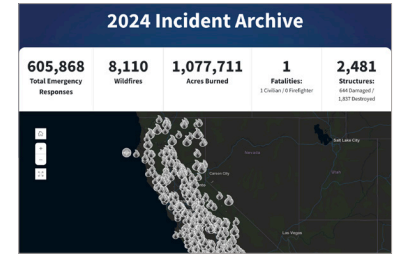

Source: Rhodium Group. Chart republished via Common Dreams, July 2024. https://www.commondreams.org/news/biden-climate-claim-not-true

Source: California Department of Forestry and Fire Protection. (2024). 2024 Incident Archive. CAL FIRE. https://www.fire.ca.gov/incidents/2024

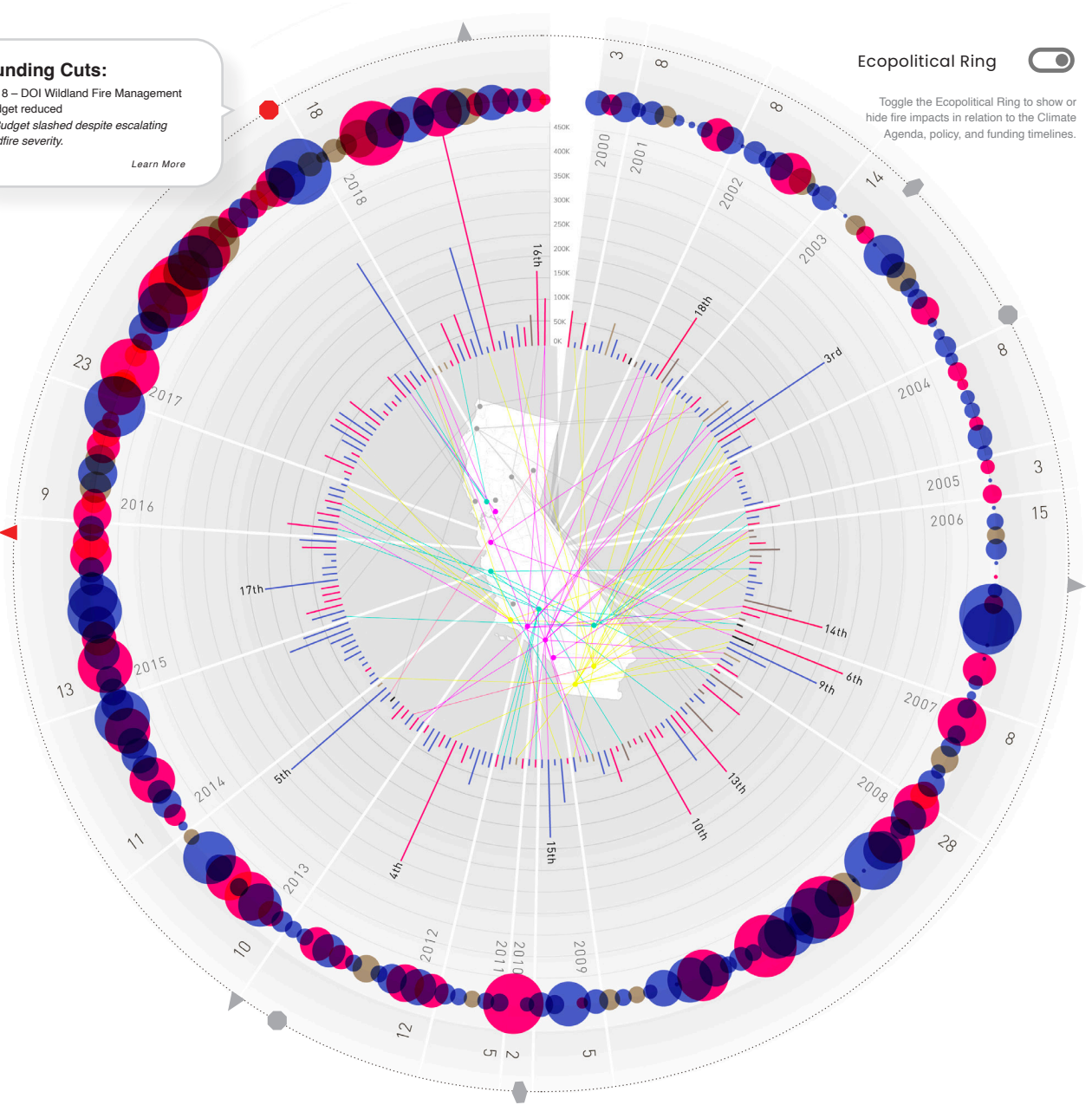

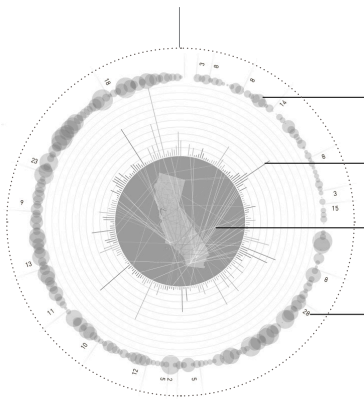

VISAP'25, Pictorials and annotated portfolios.

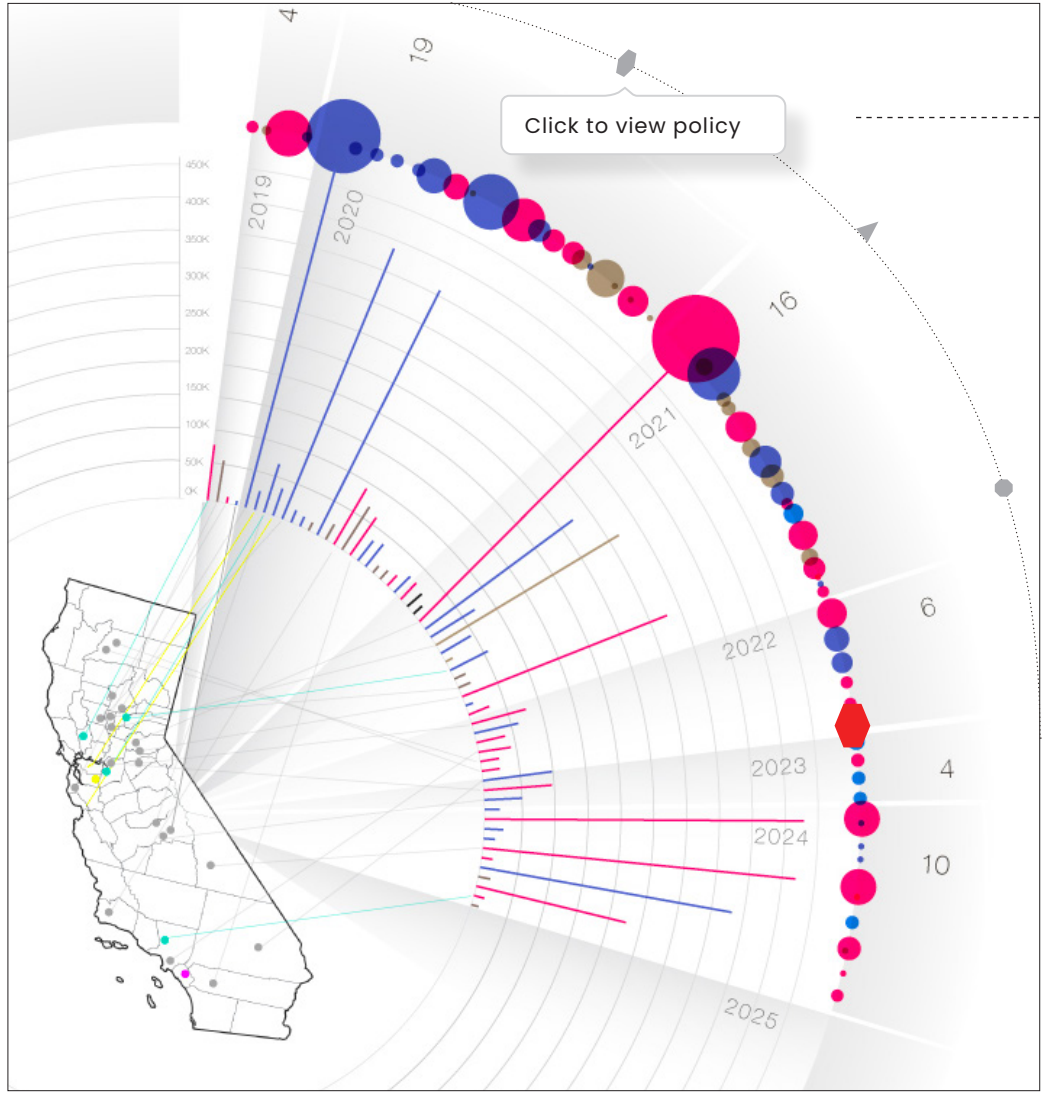
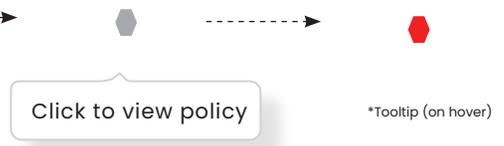

**POLICY LAYER AS DEEP INTERACTION NODE**

01: Default Icon (Pre-click)

02: Contextual Snapshots:

03: Deeper Inquiries:

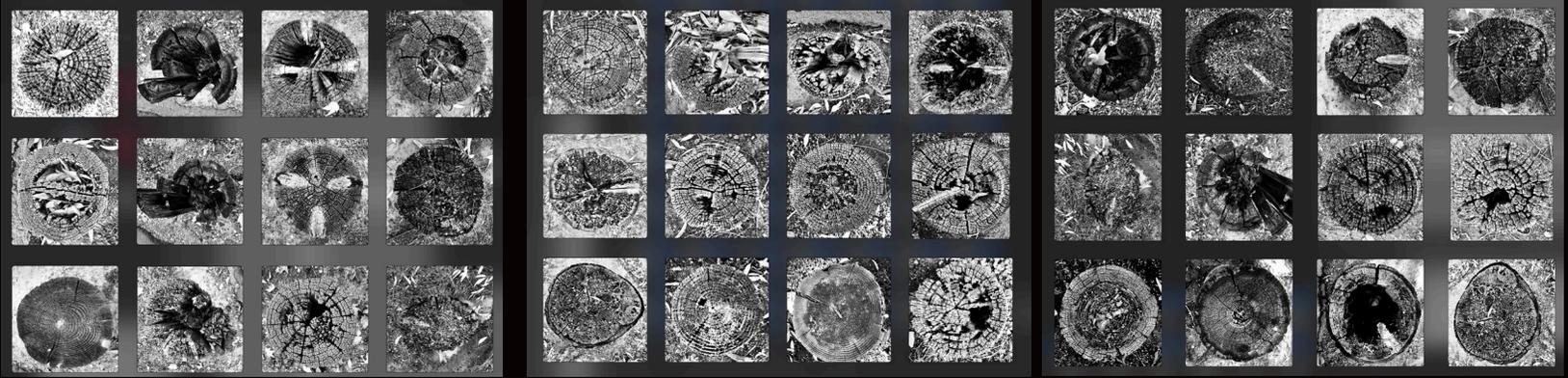

These images simulate a series of deformed tree-ring cross-sections—dynamic time capsules fractured and warped by wildfire events. Using algorithmic noise parameters, each ring becomes a visual record of ecological stress shaped by climate pressure, human actions, and infrastructural vulnerability.

Red and blue particles encode human- and natural-caused fires, layering rupture, accumulation, and resonance into a structure of event-driven memory. The top grid juxtaposes real-world fire-scarred rings with computationally generated simulations, blurring the boundary between biological memory and algorithmic storytelling.

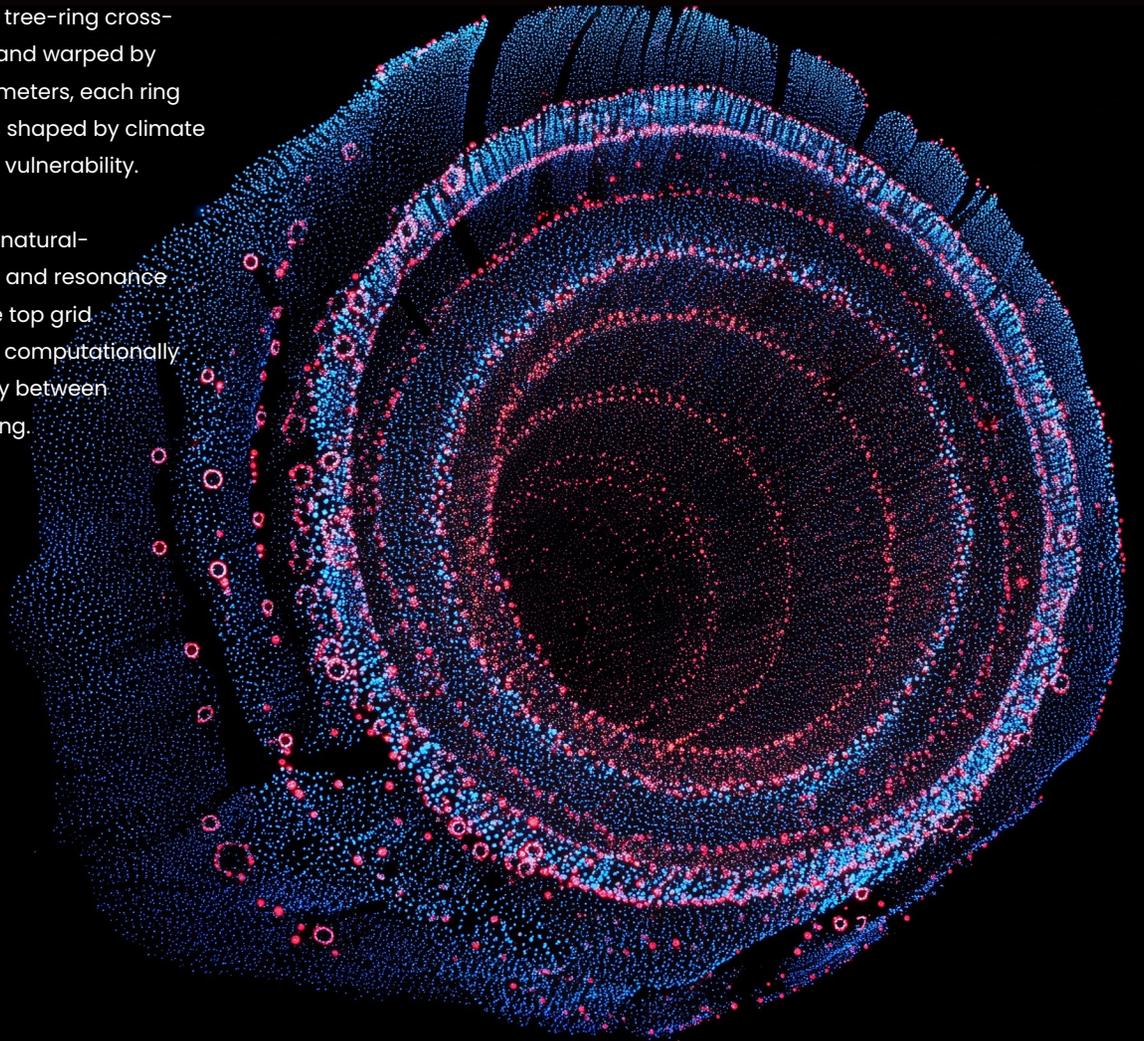

These rings are no longer static diagrams, but sensitive, responsive living archives. Designed for immersive, large-scale projection environments, they react in real time to the viewer's movement and touch—reminding us that perception itself is participation, and that data morphs through relational encounter. Every fracture is an imprint of consequence; every distortion, a reverberation of collective memory.

VISAP'25, Pictorials and annotated portfolios.

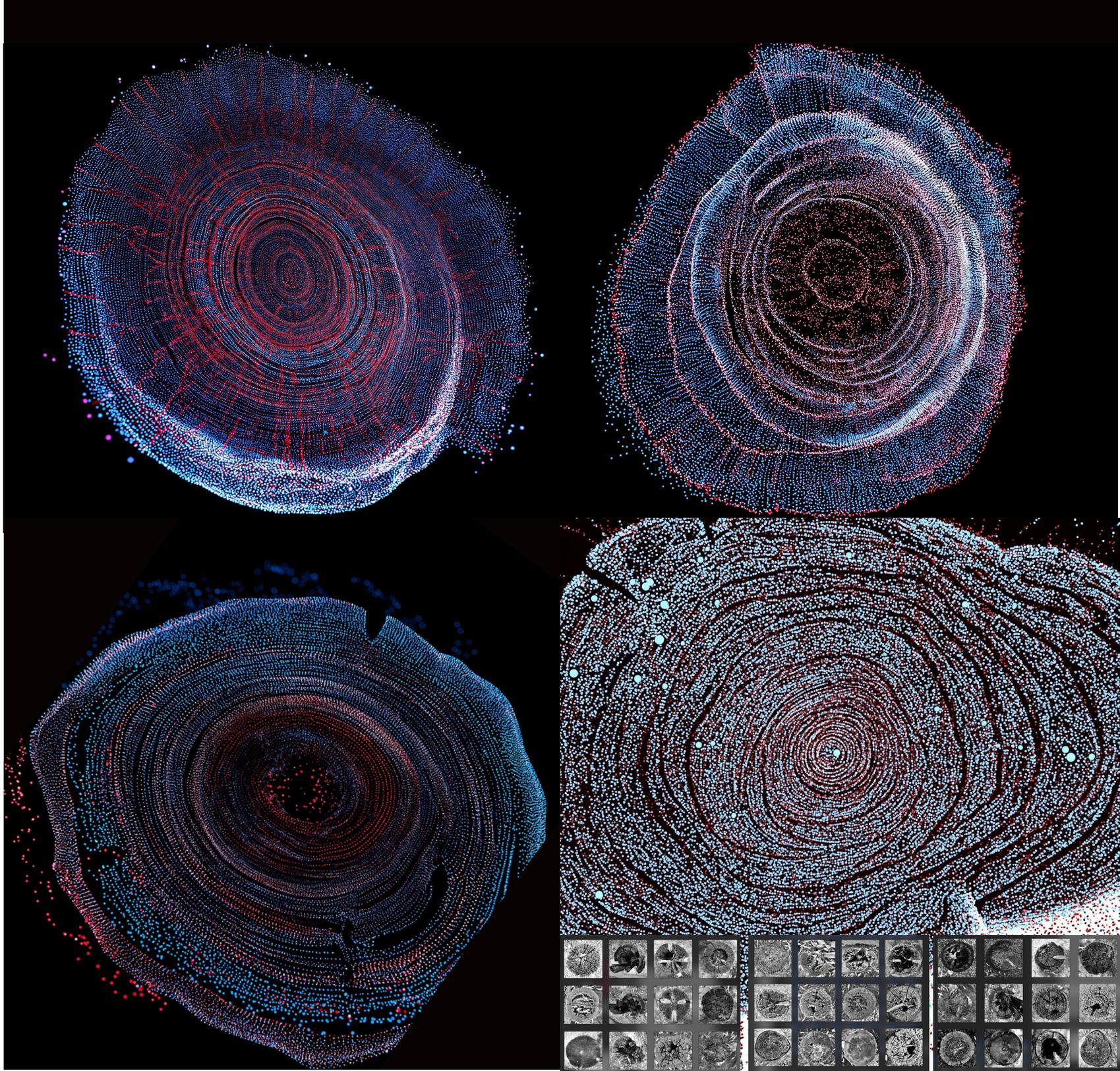

**PART 02 INSPIRATION**

The pinecone serves as both form and metaphor in this part. As a physical object, its layered spiral geometry offers a compelling structure for encoding wildfire data—visually organizing ignition causes, intensity, and temporal sequences into an intuitive, clustered interface.

Beyond its structural clarity, the pinecone is also a symbolic and emotional signifier. In many fire-adapted ecosystems, cones require heat to release seeds. This regenerative logic resonates with the project's core idea: that fire is not only destruction, but also transformation. The scorched cone becomes a site of memory, renewal, and cultural entanglement.

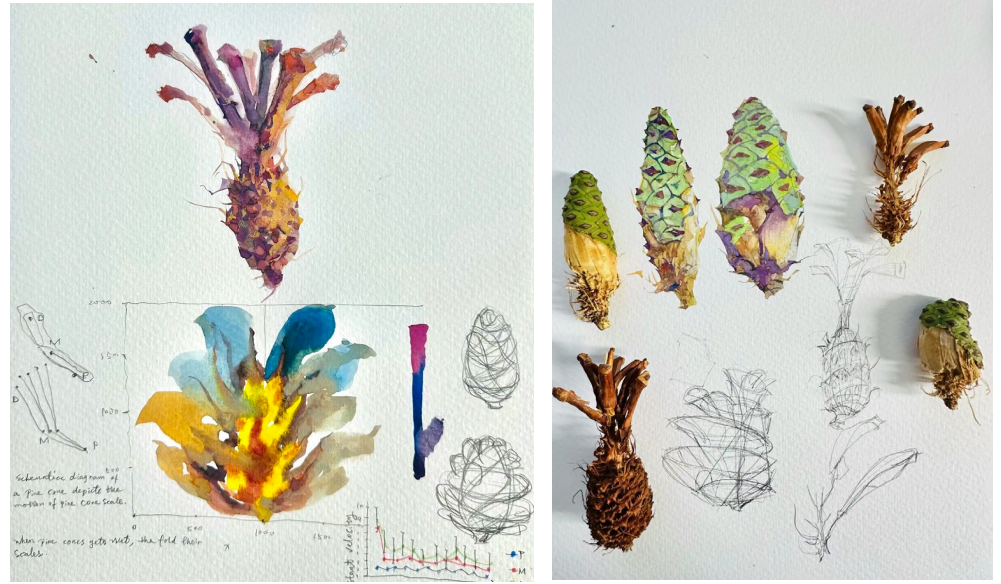

From early hand sketches and watercolor studies to 3D interactive prototypes, this section traces how the pinecone evolved from botanical reference to computational archive—bridging ecology, data, and design.

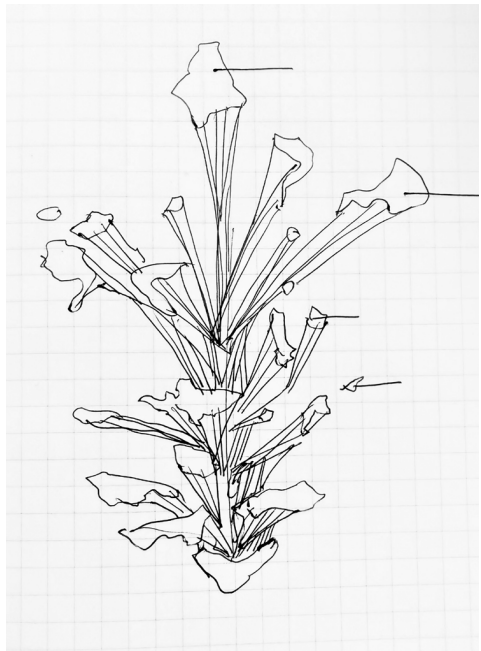
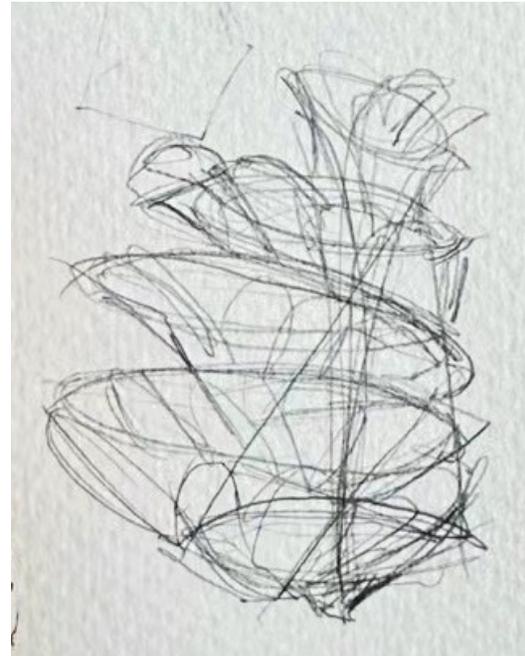
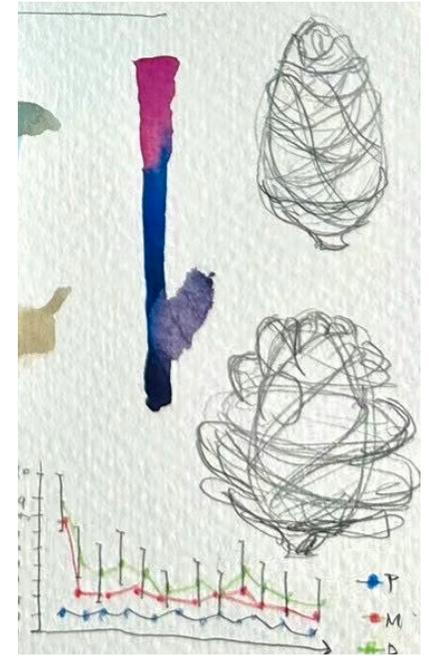

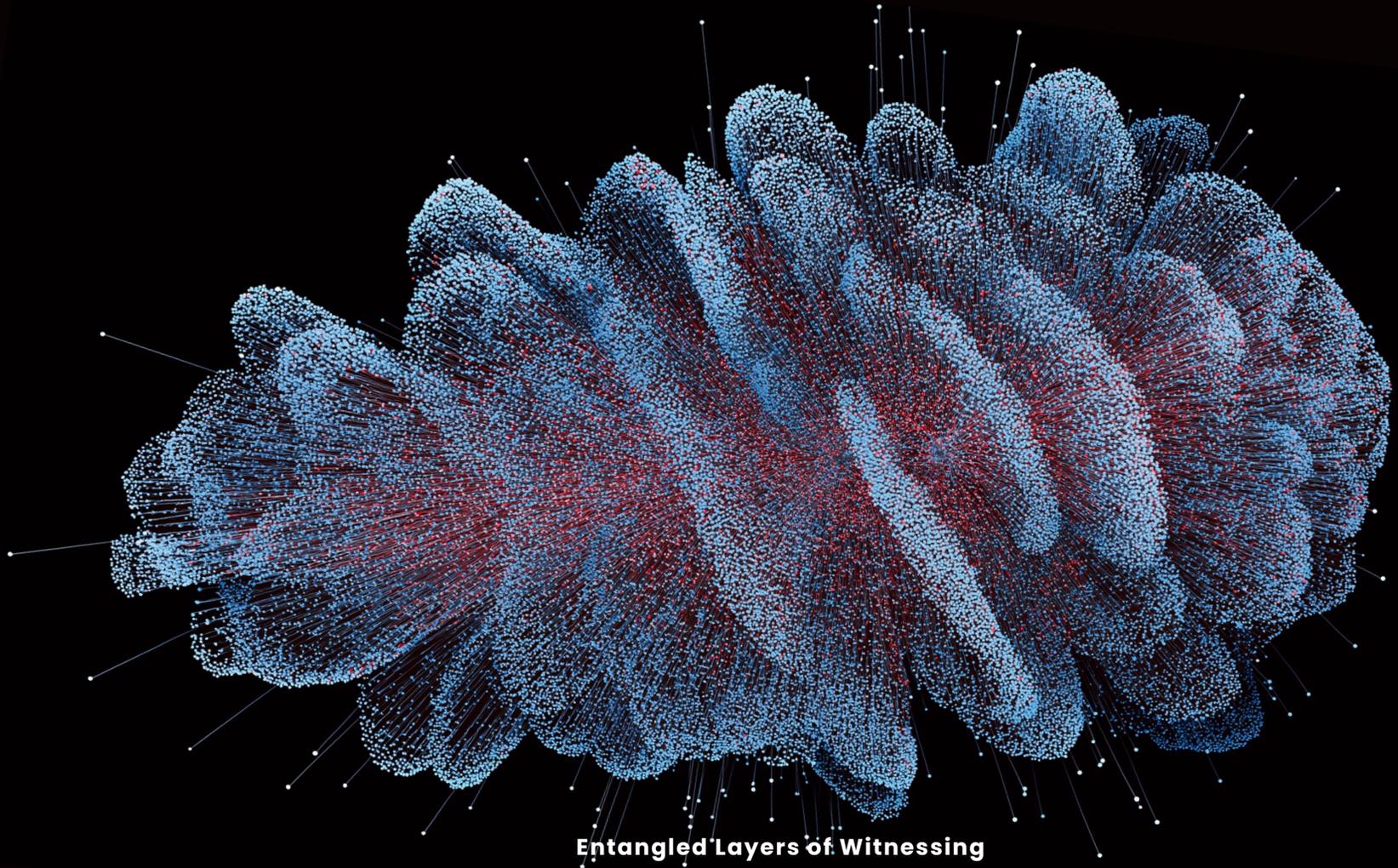

**Entangled Layers of Witnessing**

A five-layered framework for understanding wildfire beyond metrics—spanning data, policy, voice, and co-creation.

The pinecone is a living archive of wildfire memory [9]. Each scale holds a clustered dataset from a single fire—ecological facts, policies, hidden histories, personal accounts, and co-created inputs. In AR/VR, zooming into a scale opens a five-layer interface that reveals the intertwined forces behind the fire.

| Ecological Data | Policy & Infrastructure | Counter-Data | Voice Discrepancy | Participatory Layer |
|---|---|---|---|---|
| Ignition, scale, duration, and geography | Laws, timelines, utility records, regulatory events | Field research, Indigenous knowledge, missing records | Mainstream vs. local testimonies | Uploaded voices, reflections, and responses |

# Narrative as Interface

The archive unfolds across three levels: **Pinecone Overview**, **Event Archive**, and **Memory Layers**. Visitors move freely between levels, with content revealed progressively. In the archive, each ring corresponds to one memory dimension: **Time, Place, Voice, Emotion,** and **Prompt**.

### August Complex (2020)

Cause type tag "natural"
Ignition date September 4, 2020
Ignition source Lightning [7]
Fatalities 0
Evacuations 25,000 residents evacuated
Structures destroyed ~856
Firefighting cost
$193 million+
Duration 3 months (Sep-Dec)

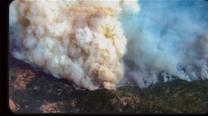
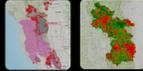

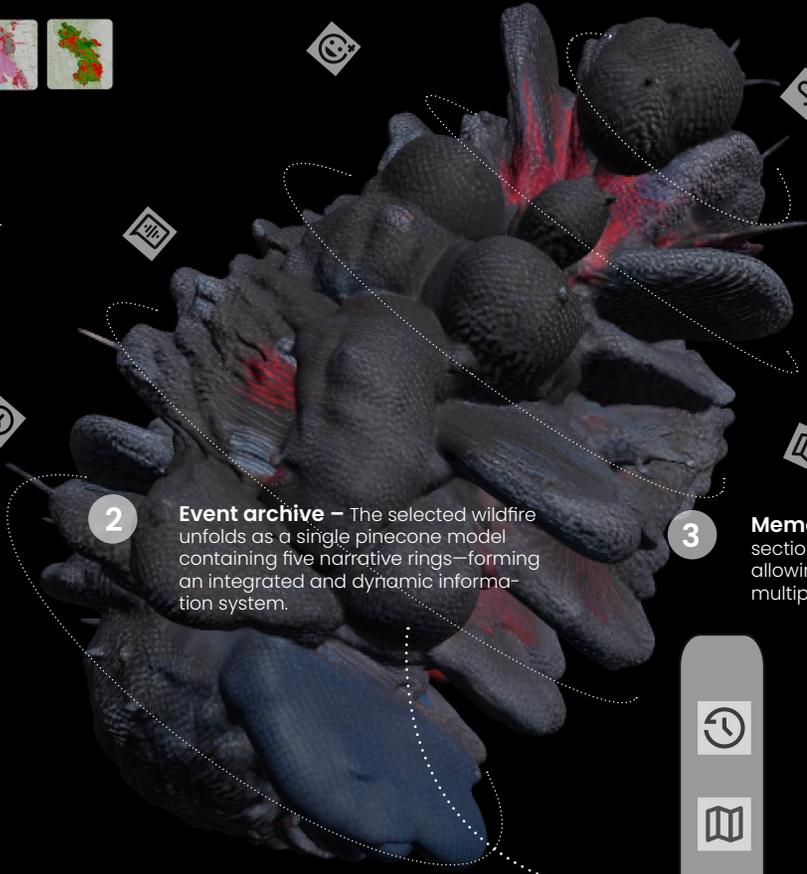

**1** **Pinecone overview** — Viewed in AR/VR, the pinecone can be rotated and zoomed. Each scale represents one wildfire event and can be selected to open its record.

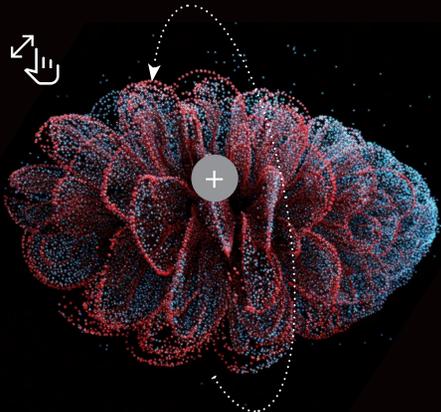

**2** **Event archive** — The selected wildfire unfolds as a single pinecone model containing five narrative rings—forming an integrated and dynamic information system.

**3** **Memory layers** — Each record expands into five sections—Time, Place, Voice, Emotion, and Prompt—allowing users to explore the wildfire event through multiple dimensions of memory and perspective.

**Time** — key dates and phases: ignition, spread, containment, recovery.

**Place** — interactive map of burn area, terrain, and affected communities.

**Voice** — testimonies from residents, responders, or officials.

**Emotion** — visual or audio signals linked to emotional responses.

**Prompt** — questions designed to encourage visitor reflection.

## Voices of Fire:

Embodied Narratives in the Pinecone Interface, Voices are collected via field interviews, community submissions, and AI-based synthesis of testimony. Each voice is visually marked with a glyph: its tone, emotion, and origin

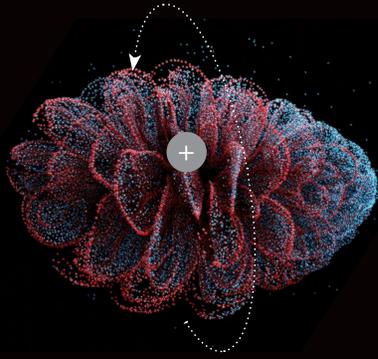

This layer amplifies the tensions between official narratives and lived experience. Viewers engage with embedded testimonies—some gathered through fieldwork, others simulated through AI—to witness how fire is remembered, narrated, or silenced.

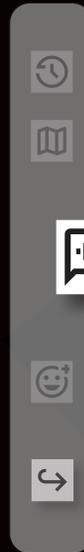

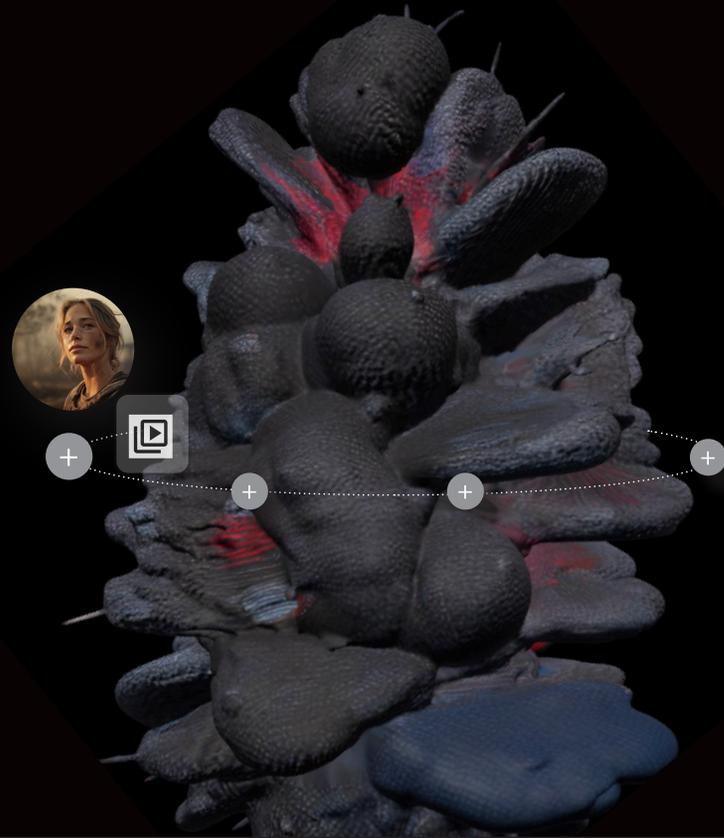

| LOSS & RECOVERY | CRISIS RESPONSE | HERITAGE PROTECTION | GOVERNANCE & JUSTIFICATION |
|---|---|---|---|
| 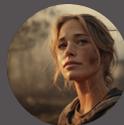 | 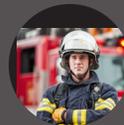 | 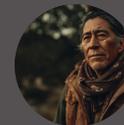 | 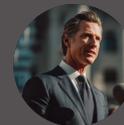 |
| Displaced residents | First responders | Indigenous land stewards | Policymakers and officials |

These voices form the foundation for an evolving co-created archive—an invitation extended in the next layer.

## Participatory Layer

Embedding visitor voices into a growing memory system of wildfire.

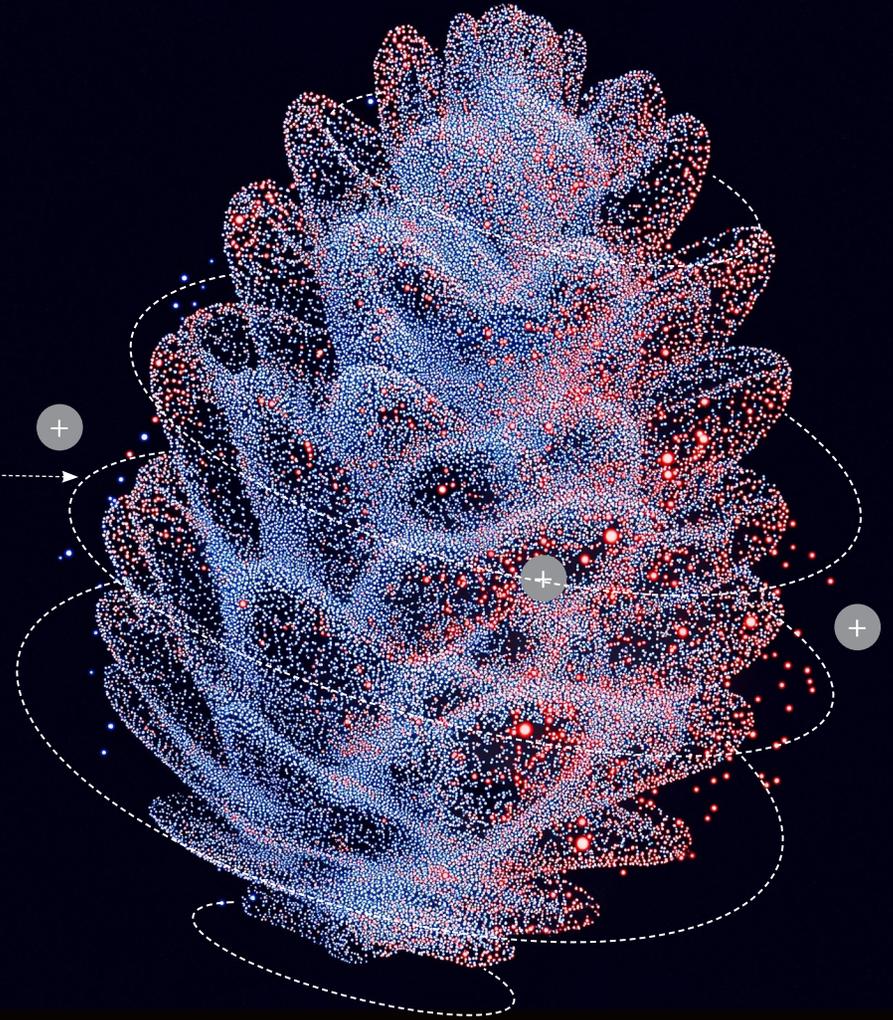

This layer turns witnessing into contribution. Visitors can upload personal reflections, speculative responses, or sensory media (audio, video, image, text), which become embedded within pinecone scales—transforming the archive into a co-authored system of care [5].

Uploaded voices will be curated and embedded into pinecone scales, forming part of future public-facing or immersive iterations.

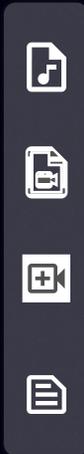

## Design Methodology & Process: From Botanical Form to Data Ritual

The Fire We Share employs an interdisciplinary methodology integrating ecological visualization, speculative narrative, and relational data design. The process unfolds in four interwoven phases:

### Phase 1: Ecological Data Framework

We compiled wildfire data (CAL FIRE, OEHHA; 2000–2024) structured by:

- **Ignition Source:** Human-caused, natural, undetermined
- **Scale:** Acres burned, duration, intensity
- **Spatial Impact:** Wildland–Urban Interface (WUI) density, regional variation
- **Case Studies:** All fires over 10,000 acres

This framework supports multi-scalar, time-based ecological analysis..

### Phase 2: Material Metaphor & Visual Encoding

We adopted the pinecone as a symbolic and structural vessel—its spiral geometry encoding time, rupture, and regeneration.

Key encoding strategies include:
**Color:** Human-caused (red) vs. natural fires (blue)
**Geometry**: Deformation visualizes ecological stress and system failure
**Scale & Density:** Encodes severity and sociopolitical exposure
**Rings of Policy:** Ecopolitical overlays mark climate actions, funding cuts, and regulatory shifts
The result is a tactile, emotionally resonant data language.

### Phase 3: Interactive Interface & Narrative Design

The pinecone becomes an interactive interface activated through embodied gestures—each designed as a ritual act of care and remembrance:

- **Touch:** Triggers maps, audio, statistics, and embedded testimonies
- **Rotate (in AR):** Invokes the act of circling a fire, sharing stories
- **Long-press:** Simulates lighting a memory flame
- **Blow (via mic input):** Spreads ember across linked events
- **Dual-touch:** Opens relational narratives between regions

Interaction becomes ignition—more than access, it's activation.

### + Community Voice Portal
**From Interaction to Contribution**
Visitors are invited to co-author the archive by uploading memories, reflections, or speculative messages (text, audio, video). These contributions are curated and embedded into the pinecone, transforming it into a polyphonic structure of care and co-witnessing.

### Phase 4: Ethical & Critical Framework

Grounded in care ethics and situated knowledge, our method treats data as relational, embodied, and politically charged. Visualization becomes a space of resonance, not resolution.

**Multiplicity & Polyphony:** Centering marginalized perspectives and divergent truths
**Care-as-Design:** Visualization as a space of resonance, contradiction, and refusal of closure
**Temporal Critique:** Revealing institutional delay, ecological trauma, and misaligned policy response

Through this lens, The Fire We Share becomes not a diagram, but a relational intervention.

## Ritual Interface Design mechanism:
### Ritual Acts of Engagement

This ritual chain of interactions—01: from Breath of Awakening to 04: Voices in the Ember—guides users through an immersive journey of memory and narrative. It evokes personal recollection and fosters emotional and sensory connection through atmospheric immersion, with the intent of sparking critical reflection and empathy.

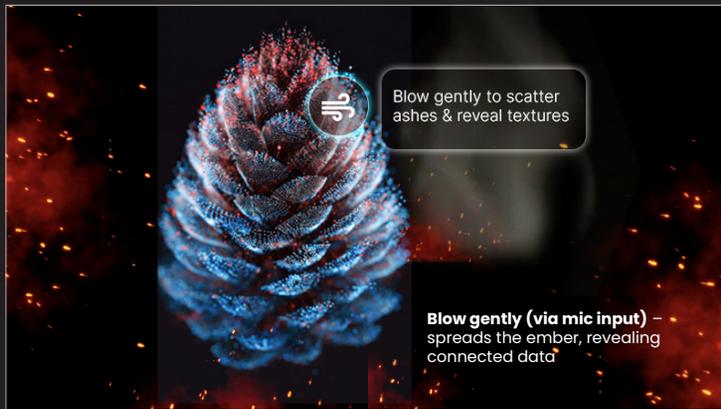

### 01: Breath of Awakening

**Interaction:** Blow to scatter ashes → reveal textures
**Effect:** AR/VR immersion in ashes and embers, triggering memory and setting mood.
**Motivation:** Breath awakens data and emotion, preparing for the story.

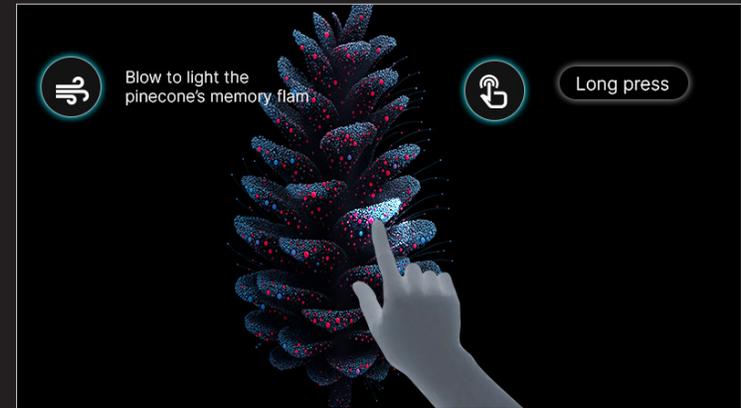

### 02: Lighting the Memory Flame

**Interaction:** Blow/Long-press → ignite memory flame
**Effect:** Flame ignites, revealing a specific wildfire event.
**Motivation:** Symbolic ignition to initiate narrative immersion.

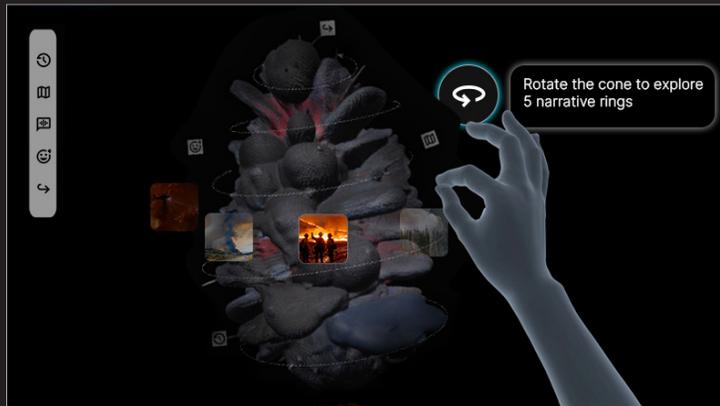

### 03: Circling the Fire

**Interaction:** Rotate → explore narrative rings
**Effect:** Five rings appear (Time, Place, Voice, Emotion, Prompt); tap to view content.
**Motivation:** Merge the metaphor of circling the fire with an accessible layered storytelling structure.

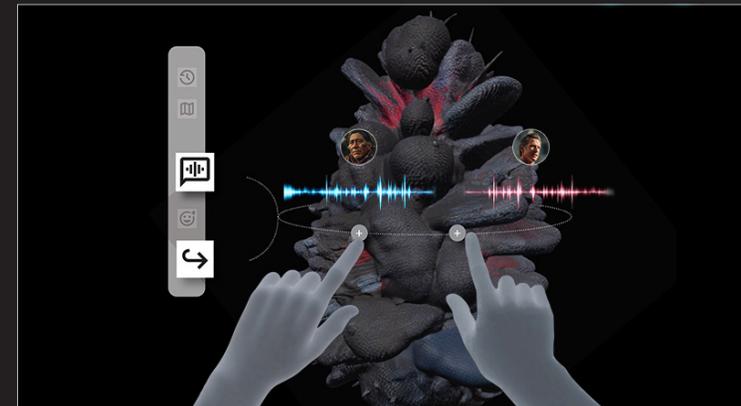

### 04: Voices in the Ember

**Interaction:** Touch voice ring → hear multiple perspectives
**Effect:** A prompt appears, followed by multiple voices from different roles.
**Motivation:** Use questions to trigger multi-perspective narrative contrasts.

## Discussion

As a California resident, I have lived through many summers punctuated by wildfire smoke, red moons, evacuation warnings, and the smell of scorched earth. In 2024, the wildfires were the most destructive in state history. By mid-2025, the fire season had already returned with familiar anxiety. It became impossible to treat fire data as something distant—as if it belonged only to scientists or statisticians.

In this pictorial, I turn the pinecone into a data structure and a ritual object—a spiral of memory, loss, and regeneration [9]. But each scale is not just a number—it is a wound, a voice, a question. The cone becomes an interface not just for viewing data, but for witnessing contradiction: **what burns, who speaks, and who is silenced.**

The pinecone is no longer symmetrical. Its fractures encode both ecological trauma and social failure. It responds not only to lightning, but to policy neglect, underfunded communities, and discarded cigarettes [8]. Through each scale, users access layered stories—mapped across time, geography, emotion, and positional truth.

**But it does not end there.** Viewers are invited to add their own voices through a Community Upload Portal—transforming the interface into a living archive of grief, resistance, and care. What began as a visualization becomes a polyphonic memory ritual.

As the curators of VISAP'25 remind us, data is not a static artifact, but a living archive. This archive does not aim to resolve, but to resonate. It holds space—for rupture, for silence, for shared discomfort.

To visualize wildfire now is not to explain it, but to stay with it. To feel its heat not only as destruction, but as memory. This is where design becomes care: a gesture not only of representation, but of co-presence and co-grieving.